\documentclass[aps,reprint,twocolumn,a4paper,superscriptaddress]{revtex4-2}

\usepackage{graphicx}
\usepackage{dcolumn}
\usepackage{bm}
\usepackage{setspace}
\usepackage{amsmath}

\begin{document}

\singlespacing

\title{Ultrafast lattice and electron dynamics induced in a PbSe crystal by an intense terahertz pulse}

\author{A. A. Melnikov}
\email{melnikov@isan.troitsk.ru}
\affiliation {Institute for Spectroscopy RAS, Troitsk, Moscow, 108840 Russia}
\author{Yu. G. Selivanov}
\affiliation {P. N. Lebedev Physical Institute RAS, Moscow, 119991 Russia}
\author{D. G. Poydashev}
\affiliation {Institute for Spectroscopy RAS, Troitsk, Moscow, 108840 Russia}
\author{S. V. Chekalin}
\affiliation {Institute for Spectroscopy RAS, Troitsk, Moscow, 108840 Russia}

\begin{abstract}

We have studied the ultrafast optical response of a PbSe crystal to an intense picosecond terahertz pulse with a peak electric field strength of up to $\sim$ 500 kV/cm. The reflectivity anisotropy signal contains oscillations at the fundamental frequency of the resonant infrared-active phonon mode as well as its second, third, and fourth harmonics. The effect is ascribed to coherent anharmonic phonons resonantly excited by the strong terahertz field. Pump terahertz pulses also induce an almost instantaneous Kerr effect and a long-lived optical anisotropy of the crystal with a characteristic decay time of $\gtrsim$ 100 ps. We consider lattice distortion and phonon-assisted side valley population as possible origins of this metastable state.

\end{abstract}

\maketitle

\section{Introduction}

Lead selenide belongs to the family of monochalcogenides of lead, tin, and germanium with the formula $MX$, where $M$ = Ge, Sn, Pb, and $X$ = S, Se, Te. These solids possess a unique set of physical properties, among which is their very high thermoelectric efficiency caused by a favorable combination of relatively high carrier mobility and low thermal conductivity \cite{He, Wei, Hu, Shafique, Filanovich, Starkov}. The latter, in turn, is a result of the peculiar lattice anharmonicity in these chalcogenides, whose mechanisms are actively studied \cite{Delaire, Jiang, Li1, Lanigan-Atkins, Skelton, Kimber}. Unusual anharmonic lattice effects observed in the $MX$ crystals are often associated with the stereochemical activity of lone electron pairs of the cation $M$ \cite{Waghmare, Walsh, Nielsen, Sangiorgio}. In cubic crystals these lone pairs induce shifts of cation atoms relative to their high-symmetry positions leading to the formation of fluctuating local dipoles and to lattice instability \cite{Jiang, Sangiorgio, Božin, Jensen}. The specific character of interatomic bonds in $MX$ monochalcogenides also causes considerable anharmonicity of their transverse optical (TO) phonon modes. These modes are often actually the soft modes associated with a structural displacive phase transition \cite{Iizumi, Wang, Huang, Shi}. Highly anharmonic TO phonons effectively scatter acoustic phonons, which carry heat, thereby lowering the thermal conductivity \cite{Delaire}.

Transverse optical phonon modes in $MX$ crystals have rather low frequencies in the terahertz range and are infrared-active. Now there exist nonlinear optical methods that allow efficient generation of picosecond terahertz pulses with high peak electric fields \cite{Fülöp, Rumiantsev}. Thus, it would be interesting to study ultrafast lattice dynamics induced via intense resonant excitation of coherent TO phonons.

Resonant pumping of infrared-active vibrations in solids by ultrashort laser pulses has emerged recently as a unique method for the ultrafast control of quantum materials and investigation of unusual nonequilibrium states \cite{Kampfrath, Nicoletti, Basov, Hübener}. A number of interesting concepts have already been demonstrated or suggested. In particular, it was shown that dipole-active atomic vibrations can couple quadratically to the electron density of a crystal, potentially enabling light-induced electronic phase transitions and superconductivity \cite{Gierz, Kennes, Sentef, Mitrano, Rini, Esposito}. Anharmonic interactions of a highly-excited phonon mode can cause nonlinear lattice distortions and induce effective magnetic fields, providing a means of controlling the structural, ferroelectric, and magnetic order \cite{Först, Subedi, Juraschek1, Hoegen, Fechner, Li2, Nova, Melnikov1, Melnikov2}.

In our recent work we used intense terahertz pulses for resonant excitation of the soft TO phonon mode of a PbTe crystal and detected nonlinear atomic oscillations at frequencies up to the third harmonic of the fundamental frequency \cite{Melnikov3}. Relying on our data, it was also possible to assume that the terahertz electric field induces transient symmetry lowering accompanied by the short-lived polar order. The lattice of PbSe is often considered to be more anharmonic in comparison to that of PbTe \cite{Tian}. Moreover, Pb-Se bonds have more “ionic” character, and, therefore, can demonstrate stronger coupling to the resonant terahertz radiation. In accordance with these assumptions, in the present work we have found that terahertz-driven anharmonic dynamics are more pronounced in PbSe and up to four harmonics of the TO phonon mode can be detected, which implies universality of these nonlinear lattice effects for lead monochalcogenides. Additionally, terahertz pulses induce a long-lived anisotropy of reflectivity of the PbSe crystal that we ascribe either to a metastable lattice distortion or to charge carriers transferred to the side valleys upon terahertz excitation. We have also detected a pronounced instantaneous terahertz electro-optic Kerr effect associated with a transient anisotropy of the electron distribution.

\section{Experimental details}

In our experiments we used nearly single-cycle picosecond terahertz pulses generated via optical rectification of femtosecond laser pulses with tilted fronts in a crystal of lithium niobate \cite{Stepanov}. The description of the laser setup can be found in our previous publications \cite{Melnikov3}. The peak electric field of the focused $\sim$ 1 ps terahertz pulses was up to $\sim$ 500 kV/cm, while their peak frequency was near 1 THz (see Fig. 1(a) and (b)). The frequency of the infrared-active TO phonon mode in $MX$ chalcogenides often depends considerably on carrier concentration $N$. In the case of soft modes, the frequency increases with $N$ \cite{Burkhard}.  For PbTe and PbSe at room temperature it lies typically in the range of $\sim$ 0.8--1.0 THz and $\sim$ 1.3--1.5 THz, respectively \cite{Burkhard, Cochran, Alperin, Tian, Landolt-Börnstein, Chen}. The crystal of PbSe that was studied in the present work was grown by the self-selecting vapor growth method adjusting the stoichiometric composition by adding a small amount of excess Se \cite{Maier, Szczerbakow}. The crystal was cleaved along the (100) plane. The room-temperature concentration of charge carriers was relatively low, $N\sim$ 5$\cdot$10$^{18}$ cm$^{-3}$ ($p$-type). Therefore, screening effects were expected to be moderate, allowing efficient excitation by resonant terahertz pulses.

In order to detect the optical response of the PbSe crystal to the pump terahertz pulse, we used weak femtosecond probe pulses at 800 nm. The pump and the probe beams were incident onto the sample at an angle of about 8$^\circ$, while the initial polarization of the probe pulse was set to 45$^\circ$ relative to the vertical polarization of the pump terahertz pulses. Intensities of the vertical and horizontal polarization components of the reflected probe pulses $I_y$ and $I_x$ were detected using a Wollaston prism and a pair of amplified photodiodes. The measurement was repeated multiple times for open and closed pump beams (modulated by an optical chopper) and the normalized difference signal $S=1-(I^*_y/I^*_x)/(I_y/I_x)$ was calculated and then averaged (here the asterisk indicates intensities measured with the opened pump beam). For small angles of incidence $S \approx - 4\varphi$, where $\varphi$ is the rotation of polarization of the reflected probe pulse. In order to measure pump-induced changes of reflectivity, a similar procedure was performed, in which intensities of the reflected probe beam and of the reference beam were measured instead of intensities of the polarization components.

\section{Results and discussion}

\begin{figure}
\begin{center}
\includegraphics{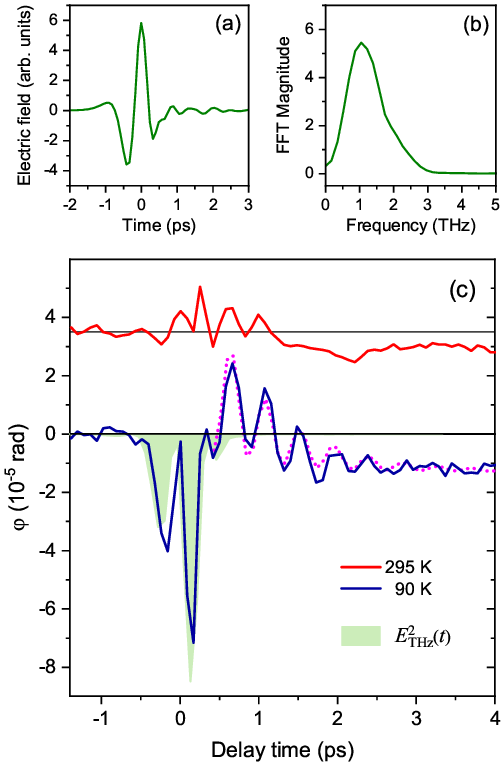}
\end{center}
\caption{\label{Fig1} (a) Temporal profile of the pump terahertz pulse. (b) Spectrum of the pulse shown in panel (a). (c) Polarization rotation $\varphi$ of the probe pulse reflected from the PbSe crystal as a function of the pump-probe delay time measured at 295 K (red solid line) and at 90 K (blue solid line). The room-temperature curve is shifted up for clarity. Normalized squared electric field of the pump terahertz pulse in arbitrary units (green area). Pink dotted line illustrates the fit by Eq. 1 (see text).}
\end{figure}

Figure 1(c) illustrates temporal evolution of the polarization rotation $\varphi$ of the probe pulse reflected from the PbSe crystal that was measured in the experiment for crystal temperatures of 295 K and 90 K. The pump terahertz pulse arrives at approximately zero delay time ($t \sim 0$) and induces anisotropy of reflectivity of the sample, which reveals itself as the nonzero $\varphi(t)$ signal. The latter consists of damped oscillations superimposed onto a monotonic transient. At lower temperatures an additional pronounced waveform appears near $t \sim 0$ that is very similar to the squared terahertz electric field. In this time window the signal $\varphi(t)$ follows $E_\mathrm{THz}^2(t)$ almost instantaneously. Performing a convolution of the $E_\mathrm{THz}^2(t)$ function with a model exponential response, it is possible to roughly estimate the characteristic timescale of the underlying physical process as $\lesssim$ 100 fs. Therefore, we ascribe this waveform to the quadratic electro-optic (Kerr) effect of the electronic origin. Moreover, we assume that this effect is determined by the intraband purely electronic processes. In this case its magnitude is expected to grow with increasing mobility of charge carriers. According to our results, at 90 K the Kerr effect is at least one order of magnitude larger than at room temperature (in fact, at 295 K it is indiscernible over the oscillations and noise). At the same time, the measured carrier mobility in our PbSe sample increases by a factor of $\sim$ 12 upon cooling to 77 K.

The observation of the terahertz Kerr effect in PbSe is not a trivial fact. Indeed, the main intraband effect of the pump terahertz pulse on charge carriers is the Joule heating of the ensemble. If the electron-electron scattering rate is sufficiently high, a thermalized electron distribution will evolve during the pulse, the temperature of which approximately follows the integral of the squared electric field (provided the characteristic relaxation time of the electron temperature is rather long). This process reveals itself as transient changes of reflectivity, which should be isotropic, since PbSe has the centrosymmetric rock-salt structure and the electron distribution function as well as the electron-electron scattering processes are expected to be symmetric. However, these isotropic changes of reflectivity are cancelled out in the process of measurement of the $\varphi(t)$ signal. Therefore, the observed terahertz Kerr effect implies a small transient anisotropy of reflectivity that probably indicates a short-lived anisotropy of the electron distribution.

This counterintuitive result could in principle be explained by invoking the concept of local structural dipoles mentioned above. The dynamic local symmetry lowering caused by the off-centering of Pb cations implies breaking of the inversion symmetry and, therefore, can enable asymmetric scattering processes. One more mechanism should be also mentioned, which requires higher-order nonlinearity of the band dispersion. It was proposed for doped semiconductors and metallic systems and is valid even for centrosymmetric crystals \cite{Ma, Almasov}. Nevertheless, the detected terahertz Kerr effect in PbSe requires further studies, which are beyond the scope of the present work. In what follows, we will analyze the signal $\varphi(t)$ measured at 90 K starting from $t\sim$ 0.5 ps, thereby excluding the Kerr waveform.

Next, we proceed to the discussion of the oscillations in the detected signal $\varphi(t)$. In femtosecond pump-probe experiments with nonmagnetic crystals, terahertz oscillations in the ultrafast optical response are usually associated with coherent atomic vibrations of certain symmetry (coherent phonons) induced by the pump pulses \cite{Dekorsy, Misochko}. For rock-salt PbSe crystals there are two phonon modes: TO mode at $\sim$ 1.3-1.5 THz and a longitudinal optical (LO) mode at $\sim$ 4 THz \cite{Landolt-Börnstein, Tian, Chen}. However, we rule out generation of coherent LO phonons of detectable amplitude by terahertz pulses. Indeed, in this case there is no resonance, while impulsive mechanisms are not effective, since the duration of the terahertz pulse ($\sim$ 1 ps) is considerably larger than the period of the LO phonon mode ($\sim$ 0.25 ps). We note that even for femtosecond pump pulses of relatively large fluence (0.5 mJ/cm$^2$ at 650 nm) we were unable to detect any oscillations in the $\varphi(t)$ response. We neglect also the sum-frequency Raman scattering of terahertz radiation \cite{Maehrline}, because in the absence of a resonant half-frequency vibrational transition this effect is expected to be very weak for the intensities of terahertz pulses used in our experiments \cite{Juraschek2, Melnikov4}. Thus, in our case we expect only coherent excitation of the infrared-active TO mode by the resonant pump terahertz pulse.

\begin{figure}
\begin{center}
\includegraphics{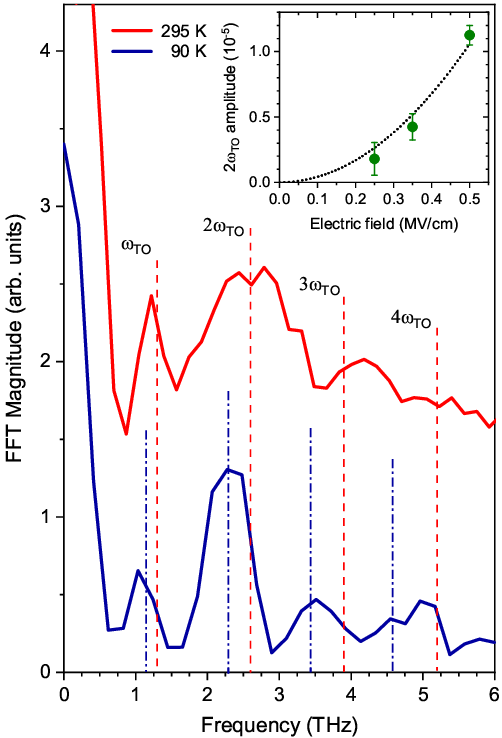}
\end{center}
\caption{\label{Fig2} Spectra of the oscillating components of the terahertz-induced signals $\varphi(t)$, measured at 295 K (red solid line) and 90 K (blue solid line). The room-temperature spectrum is shifted up for clarity. Vertical red dashed lines and blue dash-dotted lines indicate fundamental frequency of the TO mode as well as its harmonics at 295 K and 90 K, respectively. The $2\omega_\mathrm{TO}$ values are chosen as central frequencies of the most intense bands that correspond to the second harmonic. The $\omega_\mathrm{TO}$, $3\omega_\mathrm{TO}$ and $4\omega_\mathrm{TO}$ frequencies are then calculated from the $2\omega_\mathrm{TO}$ values. The inset shows the dependence of the amplitude of the second harmonic at $2\omega_\mathrm{TO}$ on the peak electric field strength of the terahertz pulse. Green circles -- experimental values, dotted curve -- a fit by a parabola $y = ax^2$.}
\end{figure}

Spectra of oscillations that are present in the $\varphi(t)$ transients measured at 295 K and 90 K are shown in Fig. 2. In both cases distinct bands appear near $\omega_\mathrm{TO}$, $2\omega_\mathrm{TO}$ and $3\omega_\mathrm{TO}$, where $\omega_\mathrm{TO}$ is the fundamental frequency of the TO phonon mode of PbSe. The band at $3\omega_\mathrm{TO}$ in the room temperature spectrum is slightly shifted to higher frequencies probably due to interference with the shoulder of the broader band at $2\omega_\mathrm{TO}$ and/or an additional small contribution caused by the Kerr effect near zero delay time. For the low temperature spectrum, the correspondence is very good. It can be seen that as the fundamental frequency decreases upon cooling due to the softening of the TO mode, the second and third harmonics shift accordingly.

We ascribe this effect to nonlinear motion of the anharmonic oscillator associated with the TO phonon mode of PbSe, which is induced via resonant coherent excitation by the intense terahertz pulse. A similar effect was observed in our previous work on PbTe \cite{Melnikov3}, however, in that case only three harmonics were detected (red-shifted due to $\sim$ 30\% lower frequency of the TO phonon mode of PbTe). For PbSe the harmonics are more pronounced already in the room temperature spectrum, while at 90 K an additional band near $4\omega_\mathrm{TO}$ appears, indicating contribution from the fourth harmonic.

An important question is how coherent infrared-active TO phonons become visible in the $\varphi(t)$ signal. Indeed, the detection scheme used in our experiments to measure $\varphi(t)$ is sensitive only to Raman active coherent phonons that cause anisotropic modulation of the refractive index of a crystal. However, for a perfect crystal with the rock-salt structure the first-order Raman scattering is forbidden \cite{Born, Ferraro}. In our previous work on PbTe we suggested that these selection rules are lifted due to the short-lived lattice distortion induced by the terahertz pulse \cite{Melnikov3}. In this case, the apparent lifetime of the oscillations should be the same as the characteristic relaxation time of the fast monotonic component of $\varphi(t)$ that is associated with this distortion. In order to check this property for PbSe we fitted the $\varphi(t)$ trace detected at 90 K by the following function (excluding the Kerr waveform):
\begin{align}
\varphi(t) &= A_1e^{(-t/\tau_1)} + A_2e^{(-t/\tau_2)}\cos(2\pi\nu t+\phi)\nonumber\\ &+ A_3(e^{(-t/\tau_3)}-1),
\end{align}
where $\nu$ is the frequency of the second harmonic $2\omega_\mathrm{TO}$ (here we include only oscillations with the highest amplitude), $\tau_1$ and $\tau_2$ are the decay times of the fast monotonic component and of the oscillations, respectively. The third term describes an additional long-lived component that appears with a delay. We have found that a satisfactory fit can be achieved if $\tau_1 \sim \tau_2 = 0.85\pm0.10$ ps and $A_1 \sim A_2$ in accordance with the above-mentioned hypothesis.

The short-lived lattice distortion can in principle be induced by the displacive force that is generated via “rectification” of the high-amplitude atomic vibrations of the TO mode due to the second-order lattice nonlinearity \cite{Först}. Then the relation $\tau_1 = \tau_2$ is naturally fulfilled. However, the mechanism of symmetry breaking for the initially centrosymmetric lattice is not clear in this case. Probably, it can be facilitated by the fluctuating local regions of lower symmetry typical for $MX$ chalcogenides. One can consider also the electric field of the pump terahertz pulse acting on these local dipoles directly and producing a short-lived macroscopic “aligned” state lacking the center of inversion.

\begin{figure}
\begin{center}
\includegraphics{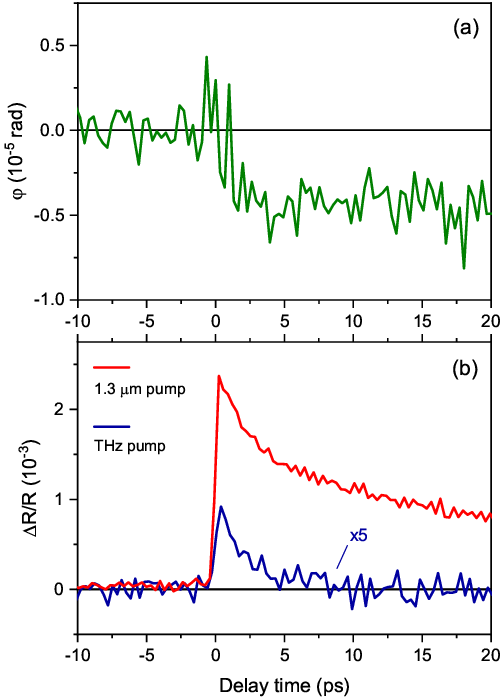}
\end{center}
\caption{\label{Fig3} (a) Temporal evolution of the polarization rotation $\varphi$ of the probe pulse in a broader time window. (b) Transient changes of reflectivity of the PbSe crystal, induced by a femtosecond laser pulse with a central wavelength of 1300 nm (top red solid line, fluence $\sim$ 0.3 mJ/cm$^2$) and by the terahertz laser pulse (blue solid line, the signal was multiplied by 5). All measurements were made at room temperature.}
\end{figure}

Besides the fast sub-picosecond monotonic component of the optical response of PbSe described by the first term in Eq. 1, we have detected a long-lived anisotropy of reflectivity that corresponds to the third term. It appears with a certain delay that is defined by the time $\tau_3$, while its decay time $T$ is so long that in our time window it can be treated as infinite (see Fig. 3(a)). Taking into account the noise, it is possible to make a rough lower estimate of this time as $T\gtrsim$ 100 ps. This long-lived anisotropy of reflectivity could be caused by a metastable distortion of the PbSe crystal lattice. However, this assumption would contradict the above interpretation, according to which the nonequilibrium lattice state of lower symmetry exists only during $\sim$ 1 ps. It is possible to consider also the electronic origin of the long-lived component. Here, we can exclude intraband processes. Indeed, as we discussed above, the anisotropy of the electron distribution that likely causes the terahertz Kerr effect has a lifetime of $\lesssim$ 100 fs. Cooling of the charge carriers heated by the terahertz pulse occurs with a characteristic time of $\sim$ 3 ps, as can be inferred from the kinetics of transient isotropic reflectivity changes (see Fig. 3(b)). Sufficiently longer decay times can be expected only in the case of interband excitation (cf. the kinetics of reflectivity changes induced by a femtosecond pulse at 1300 nm, Fig. 3(b)). However, for the terahertz pump pulses used in our experiments the probability of direct interband transitions (via multiphoton processes or tunneling) is expected to be very low taking into account the $\sim$ 0.26 eV band gap of PbSe. Nevertheless, it is possible to consider the hole transfer between $L$ and $\Sigma$ valleys, since in $p$-type PbSe the Fermi level can lie rather close to the top of the $\Sigma$ valley \cite{Herman, Xin, D'Souza, Zhu}. In this case, the population of $\Sigma$-valley states by holes could induce anisotropy of reflectivity via polarization-sensitive selection rules, if allowed by symmetry. Surprisingly, the rise time of the long-lived component $\tau_3 = 0.75\pm0.25$ ps, which is close to the decay time of the hypothetical transient lattice distortion $\tau_1$ and to the decay time of the second harmonic oscillations $\tau_2$. This fact implies that such intervalley hole transfer could be promoted by intense coherent atomic vibrations rather than by the terahertz field itself.

\section{Conclusion}

In conclusion, we have found that intense terahertz pulses generate anharmonic coherent atomic vibrations in PbSe via resonant excitation of the TO phonon mode of the crystal. We have detected the second, the third, and the fourth harmonic of the fundamental frequency in the spectrum of these oscillations. In general, nonlinear lattice dynamics in PbSe were found to be more pronounced compared to the case of PbTe that we studied earlier. Moreover, terahertz electric field induces in PbSe a nearly instantaneous electro-optic Kerr effect, indicating transient anisotropy of electron distribution. Terahertz excitation also results in a long-lived anisotropy of reflectivity of the PbSe crystal that can be associated with a metastable lattice distortion or with charge carriers promoted to the side valleys. Our results suggest universality of the observed terahertz-driven nonlinear lattice effects for lead monochalcogenides. Excitation of infrared-active TO phonons by intense terahertz pulses can be used to create unusual highly nonequilibrium lattice states in other monochalcogenides of germanium, tin, and lead in order to study lattice anharmonicity, dynamics of polar order in the presence of charge carriers, as well as the nonlinear electron-phonon interaction.

\begin{acknowledgments}

The reported study was funded by the Russian Science Foundation, Project No. 23-22-00387.

\end{acknowledgments}

\end{document}